\documentclass[twocolumn,pra,nobalancelastpage,aps,raggedbottom]{revtex4-2}
\usepackage{amsmath,amsfonts,amssymb,graphicx,hyperref,epstopdf,xcolor,tikz,scalerel,natbib}
\usepackage{mathptmx,textcomp}
\usepackage[T1]{fontenc}
\usepackage[utf8]{inputenc}

\definecolor{lime}{HTML}{A6CE39}
\DeclareRobustCommand{\orcidicon}
{
	\begin{tikzpicture} 
	\draw[lime, fill=lime] (0,0) circle [radius=0.15] node[white] {{\fontfamily{qag}\selectfont \tiny ID}};
	\draw[white, fill=white] (-0.0625,0.095) 	circle [radius=0.007];
	\end{tikzpicture}
	\hspace{-2.2mm}
}
\newcommand\orcidID[1]{\href{https://orcid.org/#1}{\orcidicon}}

\newcommand{\be}{\begin {equation}}
\newcommand{\ee}{\end {equation}}
\newcommand{\beqa}{\begin {eqnarray}}
\newcommand{\eeqa}{\end {eqnarray}}
\newcommand{\mb}{\mathbf}

\hypersetup{colorlinks,citecolor=blue,filecolor=black,linkcolor=blue,urlcolor=blue}

\begin{document}

\title{Polarization control of attosecond pulses using bi-chromatic elliptically polarized laser}

\author{Rambabu Rajpoot\orcidID{0000-0002-2196-6133}}

\author{Amol R. Holkundkar\orcidID{0000-0003-3889-0910}}
\email[E-mail: ]{amol.holkundkar@pilani.bits-pilani.ac.in}
 
\author{Jayendra N. Bandyopadhyay\orcidID{0000-0002-0825-9370}}

\affiliation{Department of Physics, Birla Institute of Technology and Science - Pilani, Rajasthan,
333031, India}

\date{\today}

\begin{abstract}

We study the higher-harmonic generation (HHG) using elliptically polarized two-color driving fields. The HHG via bi-chromatic counter-rotating laser fields is a promising source of circularly polarized ultrashort XUV radiation at the attosecond time scale. The ellipticity or the polarization of the attosecond pulses can be tweaked by modifying the emitted harmonics' ellipticity, which can be controlled by varying the driver fields. We propose a simple setup to control the polarization of the driving fields, which eventually changes the ellipticity of the attosecond pulses. A well-defined scaling law for the ellipticity of the attosecond pulse as a function of the rotation angle of the quarter-wave plate is also deduced by solving the time-dependent Schr\"odinger equation (TDSE) in two dimensions. The scaling law can further be explored to obtain the attosecond pulses of the desired degree of polarization, ranging from linear to elliptical to circular polarization.

\end{abstract}

\maketitle

\section{Introduction}
Higher-order harmonic generation (HHG) is a very promising source of coherent XUV and X-ray radiation beams with pulse duration in the attosecond regime.  The celebrated three-steps quasi-classical model describes the process of the HHG as the tunneling ionization of atomic electron followed by a free electronic motion under the driving laser field, and eventually recombination with the parent ion. Thus, emitting a harmonic photon in the transition back to the ground state \cite{Corkum1993_PRL, Lewenstein1994_PRA}. The high coherence of HHG makes it a potential spectroscopic tool for unraveling various fast processes such as delay in photo-emission \cite{Heuser2016_PRA, Schultze2010_science}, ultrafast molecular dynamics \cite{Smirnova2009_Nature}, charge migration in biologically relevant molecules, and many more, with unprecedented resolution \cite{kraus2015_Science}. Moreover, circularly polarized HHG (CP-HHG) offers unique opportunities in studying the chiral phenomena in general and has many applications in probing and characterizing the nanostructures and magnetic materials \cite{Fan2015_pnas}. The CP-HHG is a remarkable probe to study chiral-sensitive light-matter interaction dynamics, such as ultrafast spin dynamics \cite{Turgut2013_PRL, boeglin2010_Nature, radu2011_Nature, Willems2015_PRB}, x-ray magnetic circular dichroism \cite{ferre2015_NatPhoton, Bowering2001_PRL, Nahon2015_JElectronSpectrosc, Kfir2017_SciAdv, Willems2015_PRB} to name a few.

The field profile of the driving field plays a crucial role in determining the properties of the HHG. For example, the interaction of linearly polarized drivers with isotropic media generates harmonics with linear polarization \cite{Lewenstein1994_PRA}, the sinc shaped fields are recently used to control the harmonic cutoff of the HHG \cite{Rajpoot_2020}, multicolor driver fields are routinely used to enhance the efficiency and cutoff of the emitted harmonics \cite{Li2014_PRA, Khodabandeh2021_JPhysB, Greening2020_OptExp}. The ellipticity of the emitted harmonics and hence the associated attosecond pulses are useful probes to reveal the dynamical symmetries of atoms and molecules and their evolution in time \cite{Baykusheva2016_PRL, Reich2016_PRL}. However, one can not use the strongly elliptic driver for the generation of the highly elliptically polarized harmonics because the electron return to the parent ion is severely suppressed, thereby quenching the harmonic emission. On the contrary, the counter-rotating elliptically polarized driving pulses in the so-called `bicircular' configuration are reported to circumvent the electron return problem and resulted in the elliptically or circularly polarized harmonics \cite{fleischer2014_NatPhoton,Eichmann1995_PRA, Long1995_PRA}. The generated harmonics spectrum consists of pairs of left- and right-rotating harmonics.

The HHG spectrum consisting of circularly polarized harmonics with alternating helicity (the direction of rotation, clockwise or counter-clockwise) could only generate linearly polarized isolated attosecond pulses (ASPs) or pulse train, with each subsequent pulse rotated by $120^\circ$. However, the magnitude of ellipticity of ASPs can be increased if the amplitude of particular harmonics, say $3q+1$ order ($q$ is an integer), is higher than the adjacent $3q+2$ order harmonics across a range of spectral bandwidth. This can be achieved in several ways, such as by changing the intensity ratio between the two circular drivers \cite{Dorney2017_PRL, Dixit2018_PRA}, using a generating medium with a non-zero magnetic quantum number ($m$) of the ground state \cite{Milosevic2015_OptLett}, optimizing the phase-matching conditions \cite{kfir2015_NatPhoton, Kfir2016_JPhysB}, choosing different frequency ratio of bicircular fields \cite{li2017_OptQuantElectron}, and using tailored fields \cite{Ayuso2017_NewJPhys, Heslar2018_PRA}. This capability of generating higher harmonic radiation and subsequently the attosecond pulses with controlled polarization ellipticity is significant because it provides an elegant route to study response anisotropy in the matter at natural timescales. In this work, we propose a simple setup, wherein the ellipticity of the driver pulses is varied using quarter-wave plates, and the corresponding ellipticity of the generated harmonics and, in turn, attosecond pulses are tuned accordingly. A scaling law relating the ellipticity of the generated attosecond pulse with the rotation angle of quarter-waveplate is deduced by solving the TDSE in \textit{two-dimensional} Cartesian grid. The scaling law promises a very robust control over the polarization of the generated attosecond pulses by varying the rotation angles of the quarter-wave plates.  
 
The rest of the paper is organized as follows. First, in Sec. \ref{sec2}, a brief discussion of the proposed setup for pulse synthesis is given, followed by the discussion of theoretical and computational aspects of laser-atom interactions. Next, numerical results of higher harmonic generation by bi-chromatic counter-rotating driving fields are discussed in Sec. \ref{sec3} along with the polarization properties of the emitted harmonics and the generated attosecond pulses. Finally, the concluding remarks are presented in Sec. \ref{sec4}. 

\begin{figure}[t]
\includegraphics[totalheight=0.6\columnwidth]{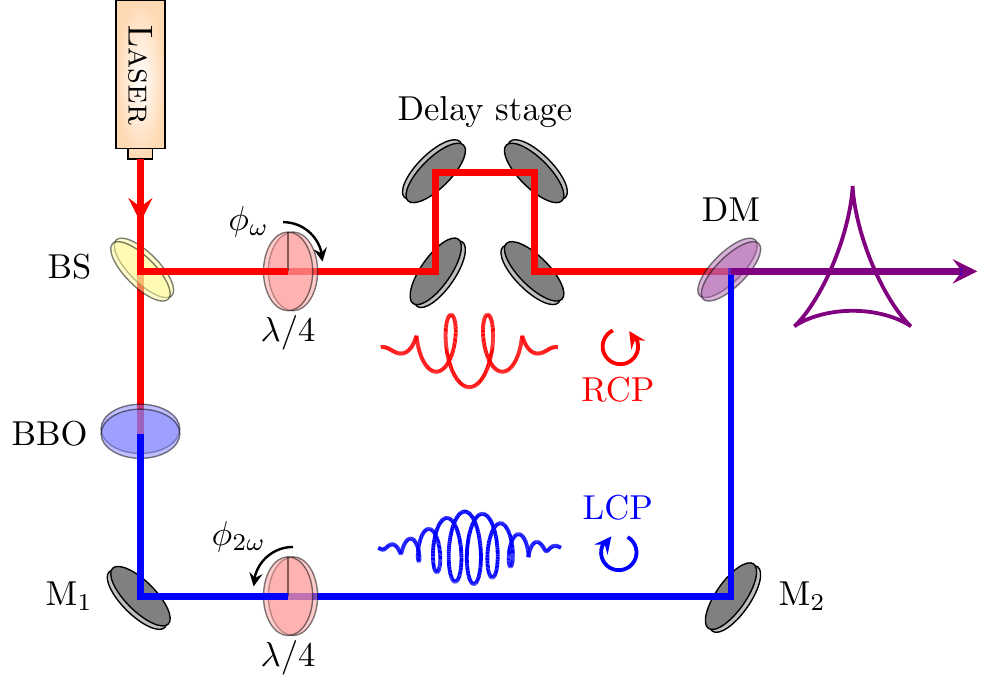}
\caption{A schematic diagram of the proposed optical setup is presented. The BBO crystal is used to generate the second harmonic ($2\omega$) of the fundamental ($\omega$) field. BS is a beam splitter, while DM is a dichroic mirror working as a beam combiner. M$_1$ and M$_2$ are perfectly reflecting mirrors. A zero-order quarter-wave plate ($\lambda/4$) is placed in each arm to control the ellipticity of the fields. When both the plates are set at, $\phi_{\omega} = \phi_{2\omega} = 45^\circ$, the two fields are counter-rotating circularly polarized, and the total electric field has a shape similar to the equilateral triangle because of the intensity ratio of $\omega$ to $2\omega$ field is 4:1.}
\label{fig_geo}
\end{figure}

\section{Numerical Methods}
\label{sec2} 

We begin this section with the description of our optical setup, followed by a brief discussion of the theoretical and computational approach adopted to calculate harmonic generation in an atomic system subject to bicircular electric fields. In order to generate two-color counter-rotating pulses, the original linearly polarized laser beam is incident onto a beam splitter (BS) as shown in Fig. \ref{fig_geo}. The BS separates the beam into two pulses with a $4:1$ intensity ratio. The weak pulse is directed onto a $\beta$-barium borate (BBO) crystal that generates the second harmonic ($2\omega$) pulse, while the pulse with higher intensity remains at the fundamental frequency ($\omega$). An achromatic zero-order quarter-wave ($\lambda/4$) plate is placed in each arm to control the ellipticity of the fields. When both the plates are rotated by $45^\circ$, the outgoing pulses are circularly polarized, with the fundamental pulse being right circularly polarized (RCP) and the second harmonic pulse being left circularly polarized (LCP). Finally, the counter-rotating pulses are combined on a dichroic mirror (DM), yielding a rosette-shaped driving electric field. By rotating one of the $\lambda/4$ plates, the polarization of the emitted harmonics and eventually of the attosecond pulses can be controlled.  Hereinafter, the atomic system of units is used, unless stated otherwise, i.e., $|e| = \hbar = m_e = 1$. 

We study the interaction of the laser pulse with a He atom by numerically solving the 2D TDSE under single-active-electron approximation. The TDSE in the length gauge is written as:
\be
	i \frac{\partial}{\partial t} \psi(\mb{r},t) = \big[- \frac{1}{2} \nabla^2 + V(r) + \mb{r}.\mb{E}(t) \big] \psi(\mb{r},t) ,
 \ee
where $\mb{E}(t)$ is the laser field and $\mb{r} \equiv (x,y)$ denotes the electron position in the two-dimensional $x$-$y$ plane.
The atomic Coulomb potential is modeled by the soft-core potential \cite{Dixit2018_PRA}:
\be
	V(\mb{r}) = - \frac{1}{\sqrt{|\mb{r}|^2 + a_0}},
	\label{pot}
 \ee
where the soft-core parameter $a_0$ is dependent on the ionization potential of the atom under study. For He-atom, $a_0 \sim 0.07$ is considered such that the ground state energy (ionization potential) of the valence 1s orbital, $E_{1s} \sim -0.904$ a.u. ($\sim$ 24.6 eV) is obtained, which is close to the experimental value of the first ionization potential of helium.
%a0 = 0.0697

The initial state is obtained by the imaginary-time propagation method \cite{Bader2013_JChemPhys}. The TDSE is propagated on a 2D Cartesian grid using the time evolution operator $U(t_0+\Delta t, t_0)$ on the initial state wavefunction $\psi_0(\mb{r},t_0)$,
\be
	\psi(\mb{r},t_0+\Delta t) \simeq U(t_0+\Delta t, t_0) \psi_0(\mb{r},t_0).
 \ee
The TDSE is solved numerically by adopting the split-operator technique \cite{Feit1982_JCP}. A mask function,
\be
	M_{abs}(r) = \frac{1}{1 + \exp[1.2 |r| - r_\text{abs}]},
 \ee
is multiplied to the $\psi(\mb{r},t)$ at each time step to avoid any nonphysical reflections at the spatial grid boundaries. The time-dependent dipole acceleration $\mb{a}(t)$ is evaluated following the Ehrenfest theorem as \cite{sandPRL_1999}:
\be
	\mb{a}(t) = - \langle \psi(\mb{r},t) | \nabla V(r) + \mb{E}(t) | \psi(\mb{r},t) \rangle.
 \ee
The harmonic spectra is then obtained by performing the Fourier transform of $\mb{a}(t)$, i.e.,
\be
	S_{\kappa}(\omega)  = \Big| \frac{1}{\sqrt{2\pi}} \int a_{\kappa}(t) e^{-i\omega t} dt \Big|^2 = \big| a_{\kappa}(\omega) \big|^2,
 \ee 
where, $\kappa$ denotes the associated $x$ or $y$ components.
To describe the polarization properties of the HHG, the intensity of the left- and right-rotating components can be obtained by
\be
	D_{\pm} = \big| a_{\pm}(\omega) \big|^2,
	\label{handedness}
 \ee
where $a_{\pm}(\omega) =  [ a_x(\omega) \pm i a_y(\omega) ]/\sqrt{2}$.	The ellipticity of the harmonics is calculated using the relation

\be
	\epsilon = \frac{ |a_+| - |a_-| }{|a_+| + |a_-|}.
	\label{ellipticiy}
 \ee

The parameter $\epsilon$ varies in the interval from $-1$ to $+1$, and the sign of $\epsilon$ defines the helicity of the harmonics. The harmonics rotating in a counter-clockwise direction have positive helicity while those rotating in a clockwise direction have negative helicity. The temporal profile of an ASP is obtained by superposing several harmonics as \cite{Liu2011_PRA}:	
\be
	I_{\kappa}(t) = \big| \sum_{q} a_{\kappa q} \exp[iq\omega t] \big|^2,
	\label{Ix}
 \ee
where $q$ is the harmonic order and $a_{\kappa q}$ represents the inverse Fourier transformation given as: 
\be
	a_{\kappa q} = \frac{1}{\sqrt{2}} \int a_{\kappa}(t) \exp[-iq\omega t] dt.
 \ee

The bicircular field is obtained by combining two counter-rotating elliptically polarized laser fields at $\lambda_1 = 800$ nm ($\omega$-field) and $\lambda_2 = 400$ nm ($2\omega$-field) wavelengths, respectively. The driving laser field in the $x$-$y$ polarized plane is defined as:

\be 
\begin{split}
\mb{E}(t) = f(t)\{ &E_1 [ \cos(\omega t + 2\phi_{\omega}) \mb{\hat{e}}_x + \cos(\omega t) \mb{\hat{e}}_y ] +\\
                   &E_2 [ \cos(2\omega t)  \mb{\hat{e}}_x + \cos(2\omega t + 2\phi_{2\omega}) \mb{\hat{e}}_y]\} 
\end{split}
\label{electric_driver}
\ee
with $0.5E_1 = E_2 \approx 0.05$ a.u. (corresponding intensity $I = 10^{14}$ W/cm$^2$). The temporal pulse envelope $f(t)$ has a trapezoidal shape with 2 cycle raising and falling edges and 5 cycle plateau (in units of $\omega$-field). The angles $\phi_{\omega}$ and $\phi_{2\omega}$ show the rotation of quarter-waveplates [refer Fig. \ref{fig_geo}], corresponding to $\omega$ and $2\omega$ fields, respectively. When both the plates are set at $\phi_{\omega} = \phi_{2\omega} = 45^\circ$, the two fields are circularly polarized and the total electric field has trefoil rosette shape. The fundamental field ($\omega$) is rotating counter-clockwise, while the second harmonic field ($2\omega$) is rotating in clockwise direction.
 
We have considered the spatial simulation domain of $\pm 150$ a.u. along both $x$ and $y$ directions. The value of parameter $r_\text{abs} = \pm 142$ a.u. is considered. The spatial step $\Delta x = \Delta y \approx 0.29$ a.u. is used and the simulation time step $\Delta t = 0.005$ a.u. is considered, which is well within the criteria $\Delta t \lesssim 0.5(\Delta x)^2$. The convergence is tested with respect to the spatial grid as well as space and time steps. Our simulation utilizes widely used \textit{Armadillo} library for linear algebra purpose \cite{Sanderson2016}.
    
\section{Results and discussions}
\label{sec3}

\begin{figure}[b]
\includegraphics[totalheight=0.5\columnwidth]{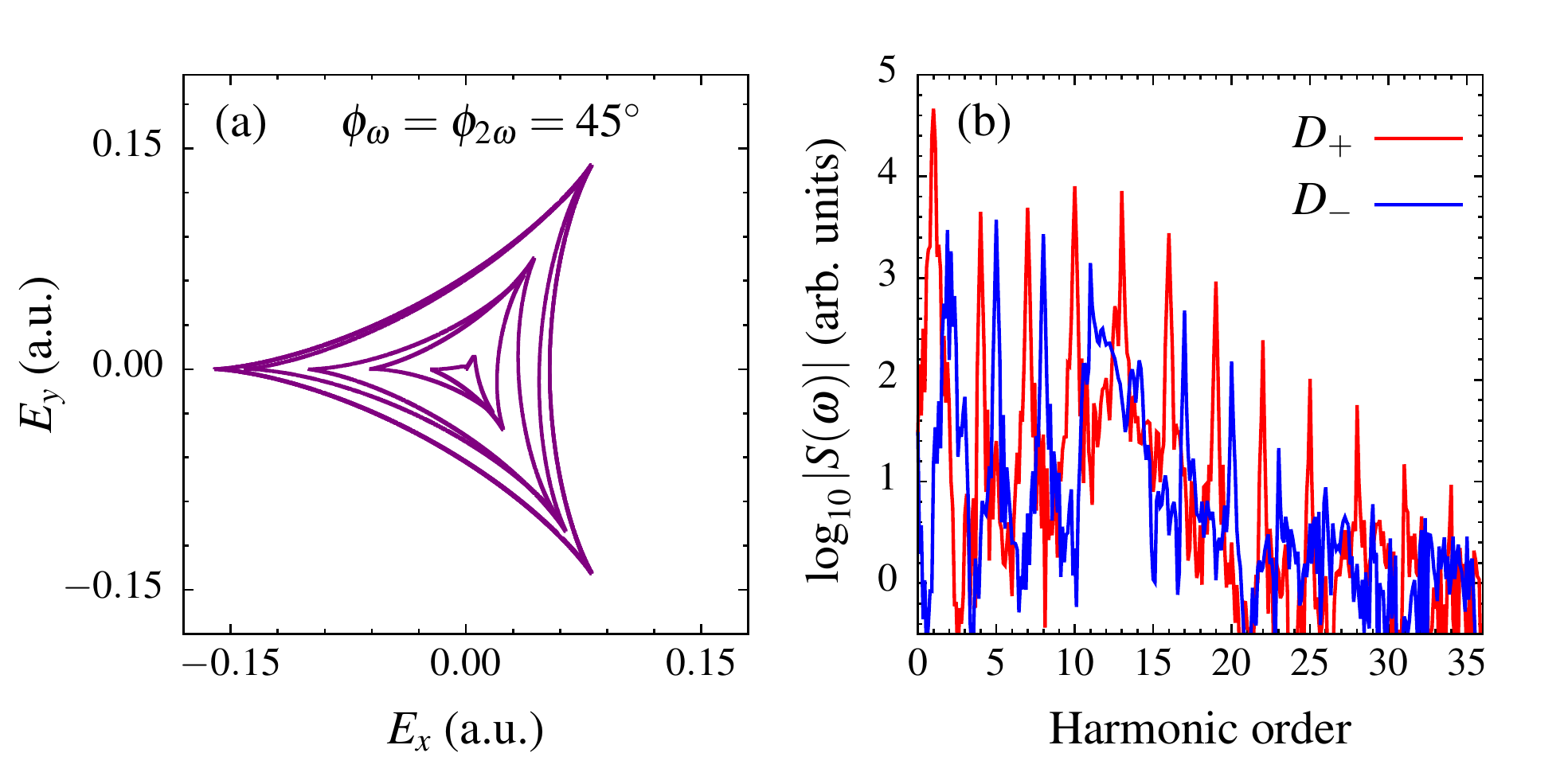}
\caption{Lissajous curve for the electric field amplitude of the bicircular counter-rotating driving field is presented in (a). The field amplitude ratio of the two drivers ($\omega : 2\omega$) is 2:1. The corresponding high-order harmonic spectrum of helium is shown in (b). The harmonic order is taken in the units of the fundamental ($\omega$) field.}
\label{hhg4545}
\end{figure} 

We start our discussion by describing the \textit{bicircular} pulse scheme and the HHG by such pulses. The scheme combines two co-planar counter-rotating circularly polarized pulses at fundamental ($\omega$) and its second harmonic ($2\omega$) frequencies. The total electric field of the two pulses traces a threefold rosette shape having symmetry with respect to a rotation of $120^\circ$. Upon interaction with the target medium, the electric field guides the tunneled-out electron away from the parent ion and back again in every \textit{one third} of the fundamental laser cycle ($T$) \cite{Lukas2015_PRL, Galan2017_OptExp}. In an isotropic and time-independent medium, this leads to a train of short XUV bursts emission, each with a linear polarization that rotates in space by $120^\circ$. In the time domain, the XUV burst is emitted every $T/3$ duration, totaling three bursts per cycle of the fundamental field. In the frequency domain, circularly polarized harmonics of order $3q+1$ and $3q+2$ are emitted and co-rotate with the fundamental and the second harmonic fields, respectively. The emission of harmonics corresponding to $3q$ orders are forbidden due to the threefold dynamical symmetry of the system \cite{Milosevic2000_PRA, fleischer2014_NatPhoton, Ivanov2014_NaturePhoton, Pisanty2014_PRA, huang2018_NatPhoton}. In Fig. \ref{hhg4545}(a), the Lissajous curve for the electric field amplitude of the bicircular driver is shown. The driving field is obtained by setting the rotation angle ($\phi_{\omega}$, $\phi_{2\omega}$) of two quarter-waveplates at $45^\circ$. As discussed for the case of circularly polarized $\omega-2\omega$ fields of equal amplitude, the total electric field traces a three-lobed structure having symmetry with respect to a rotation of $120^\circ$. In our case, the total electric field has threefold dynamical symmetry. However, due to the amplitude ratio of $2:1$ between the two fields ($\omega$, $2\omega$), the bicircular field \textit{resembles} as equilateral triangles instead of the three-lobed structure. The total field co-rotates with the $\omega$ field, i.e., counter-clockwise. In Fig. \ref{hhg4545}(b), we present the corresponding HHG spectrum consists of $3q+1$ order (red lines) right-handed and $3q+2$ order (blue lines) left-handed circularly polarized harmonics [refer Eq. \ref{handedness}]. Indeed, the $3q+1 (3q+2)$ order harmonics correspond to the absorption of a net amount of $q+1$ photons of $\omega (2\omega)$ field and a net amount of $q$ photons of $2\omega (\omega)$ field, thus co-rotates with the $\omega (2\omega)$ field to fulfill the angular momentum conservation. On the other hand, $3q$ order harmonics correspond to the absorption of a net amount of $q$ photons of each $\omega$ and  $2\omega$ fields, i.e., total $2q$ photons, preserving the parity of initial state and thus recombination by emitting a single photon of frequency $3q\omega$ is parity forbidden \cite{Galan2017_OptExp, Galan2018_PRA, Heslar2019_PRA}. It is worth mentioning that the precluded harmonics at frequency $3q\omega$ are corresponding to the harmonic peaks at the integer multiple of summary frequency $\omega + 2\omega = 3\omega$ and are not related to the third harmonic of the fundamental ($\omega$) field \cite{Andreev2020_LaserPhys}. In Fig. \ref{hhg4545}(b), the positions of harmonic peak match well with those predicted by the selection rules as mentioned earlier. However, the selection rules consider the driving field components ($\omega, 2\omega$) to be perfectly monochromatic. In the actual physical conditions, the driving field has a finite extent in time; hence the harmonic peaks have a finite width, and their polarization will vary throughout the HHG spectrum. The driving field causes the electron ionization, acceleration, and recombination thrice in an optical cycle of $\omega$ field. This dynamical symmetry of the driving field translates into the HHG spectrum, where the harmonics appear in pairs with opposite helicity and the $3q\omega$ harmonics are suppressed. It can be further seen from Fig. \ref{hhg4545}(b) that the intensities of the $(3q+1)\omega$ harmonics (red) are higher than the adjacent $(3q+2)\omega$ harmonics (blue) throughout the spectrum. As stated above, the $(3q+1)\omega$ harmonics correspond to the absorption of one extra photon of $\omega$ field, while the $(3q+2)\omega$ harmonics corresponds to the absorption of one extra photon of $2\omega$ field. By taking the field amplitude ratio $E_1:E_2 = 2:1$, the emission of $(3q+1)\omega$ harmonics is favored as can be seen in Fig. \ref{hhg4545}(b). This facet is earlier discussed and elaborated in Refs. \cite{Barth2011_PRA, Barth2013_PRA, Kaushal2016_PRA, Galan2018_PRA}. %The fundamental mechanisms responsible for the different contrast between neighboring harmonics. The difference in the contrast is because of the absorption of one extra photon of $\omega$ field of $(3q+1)\omega$ harmonics, while the $(3q+2)\omega$ harmonics correspond to the absorption of one extra photon of $2\omega$ field. By taking the field amplitude ratio $E_1:E_2 = 2:1$, the emission of $(3q+1)\omega$ harmonics is favored as can be seen in Fig. \ref{hhg4545}(b). This facet is earlier discussed, and elaborated \cite{Barth2011_PRA, Barth2013_PRA, Kaushal2016_PRA, Galan2018_PRA}.

\begin{figure}[t]
	\includegraphics[totalheight=0.95\columnwidth]{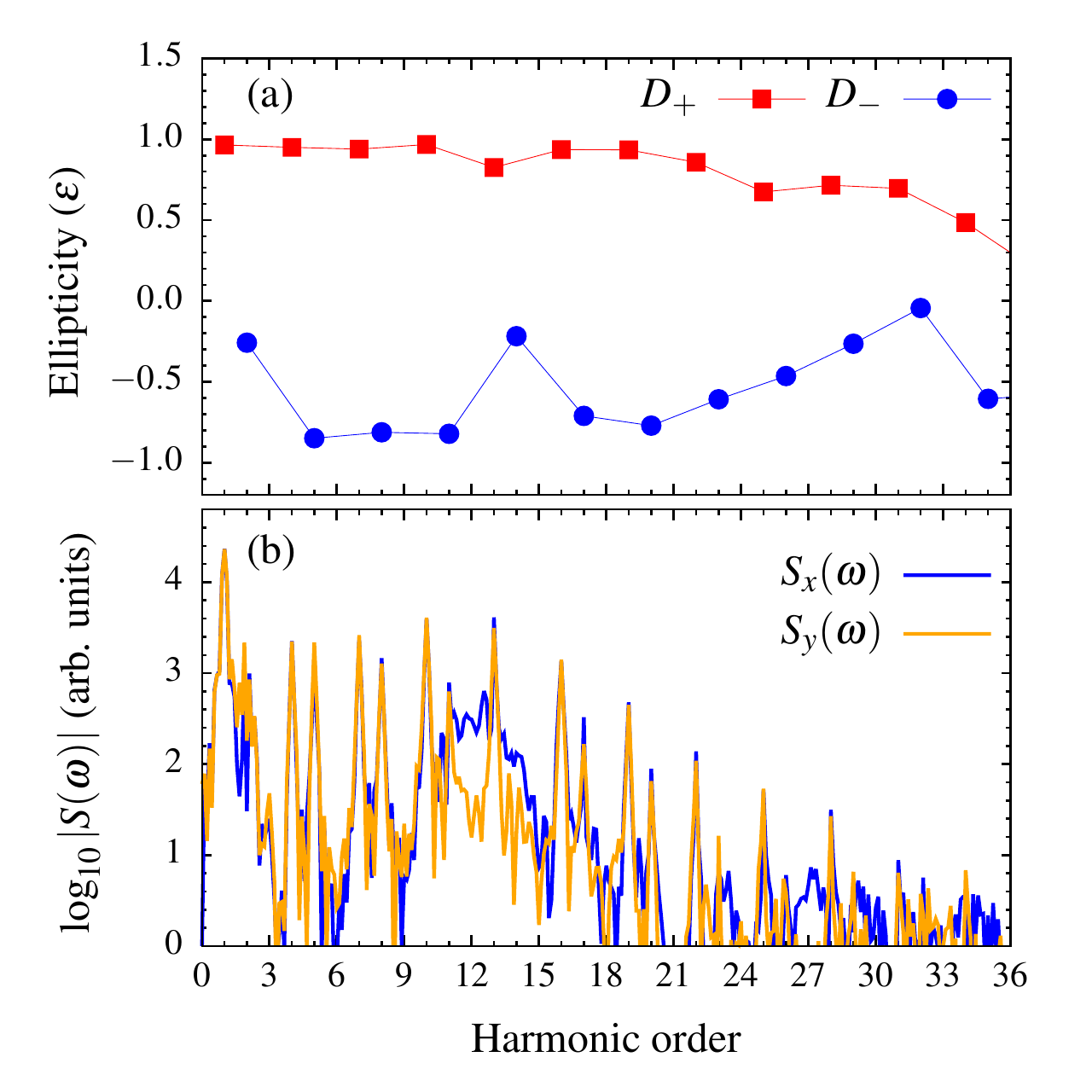}
	\caption{ (a) Ellipticity of the harmonic peaks corresponding to the HHG spectra in Fig. \ref{hhg4545}(b) are presented, and  associated $S_x$ and $S_y$ components are also shown (b). The red square curve corresponds to the harmonic component co-rotating with the $\omega$ field, while the blue circle curve represents the harmonic component co-rotating with the $2\omega$ field.}
	\label{hhg4545_eps}
\end{figure}  

Figure \ref{hhg4545_eps}(a) shows the ellipticity ($\epsilon$) of individual harmonics [refer Eq. \ref{ellipticiy}] emitted in the HHG spectrum of He atom [refer Fig. \ref{hhg4545}(b)]. The harmonics of order $3q+1$ (red square curve) have the same helicity (counter-clockwise) as the $\omega$ field, and the harmonics of order $3q+2$ (blue circle curve) have the same helicity (clockwise) as the $2\omega$ field. However, the harmonics are not exhibiting perfect circular polarization ($-1 < \epsilon < +1$), and the magnitude of the ellipticity of both types of harmonics $3q+1$ and $3q+2$ is decreasing along with the increasing harmonic order. This deviation from perfect circular polarization shows that there are some temporal asymmetries present in the system. For example, the temporal asymmetries introduced by the rising and falling edges of driving field \cite{barreau2018_NatureComm, Heslar2019_PRA}, fast ionization of generating medium \cite{barreau2018_NatureComm}, and the excitation of the bound states and subsequent near-resonant emissions \cite{Galan2017_OptExp, Heslar2018_PRA}. Generally, the perfect circular polarization is expected if the $x$ and $y$ components of the harmonic radiation, i.e., $S_x(\omega)$ and $S_y(\omega)$ have equal contributions to the total HHG spectrum. In Fig. \ref{hhg4545_eps}(b), the $x$ and $y$ components of the HHG signal are shown, and it can be seen that both the components are not well overlapped near the peaks of the harmonics, thus causing a deviation from the perfect circular polarization of the generated harmonics. Also, with increasing harmonic order, the contributions of the $S_x(\omega)$ and $S_y(\omega)$ components differ more and more near the harmonic peaks. This increasing difference alters the ellipticity and the polarization degree of the harmonics as observed in Fig. \ref{hhg4545_eps}(a). One can see that the ellipticity of the 14th harmonic is much smaller than the ellipticity of the neighboring harmonics. It is also seen in Fig. \ref{hhg4545_eps}(b) that the intensity of this harmonic is greatly suppressed in the HHG spectrum. This suppression in harmonic intensity suggests a few mechanisms responsible for the formation of the 14th harmonic and needs further investigation. One possible factor could be the applied intensities of the fundamental ($\omega$) and the second harmonic ($2\omega$) driving fields.  No such suppression in the intensity of the 14th harmonic is earlier reported in the HHG spectrum of helium \cite{Galan2018_PRA, Dixit2018_PRA}, though the applied intensities of $\omega-2\omega$ fields are different in their cases.

\begin{figure}[b]
	\includegraphics[totalheight=0.95\columnwidth]{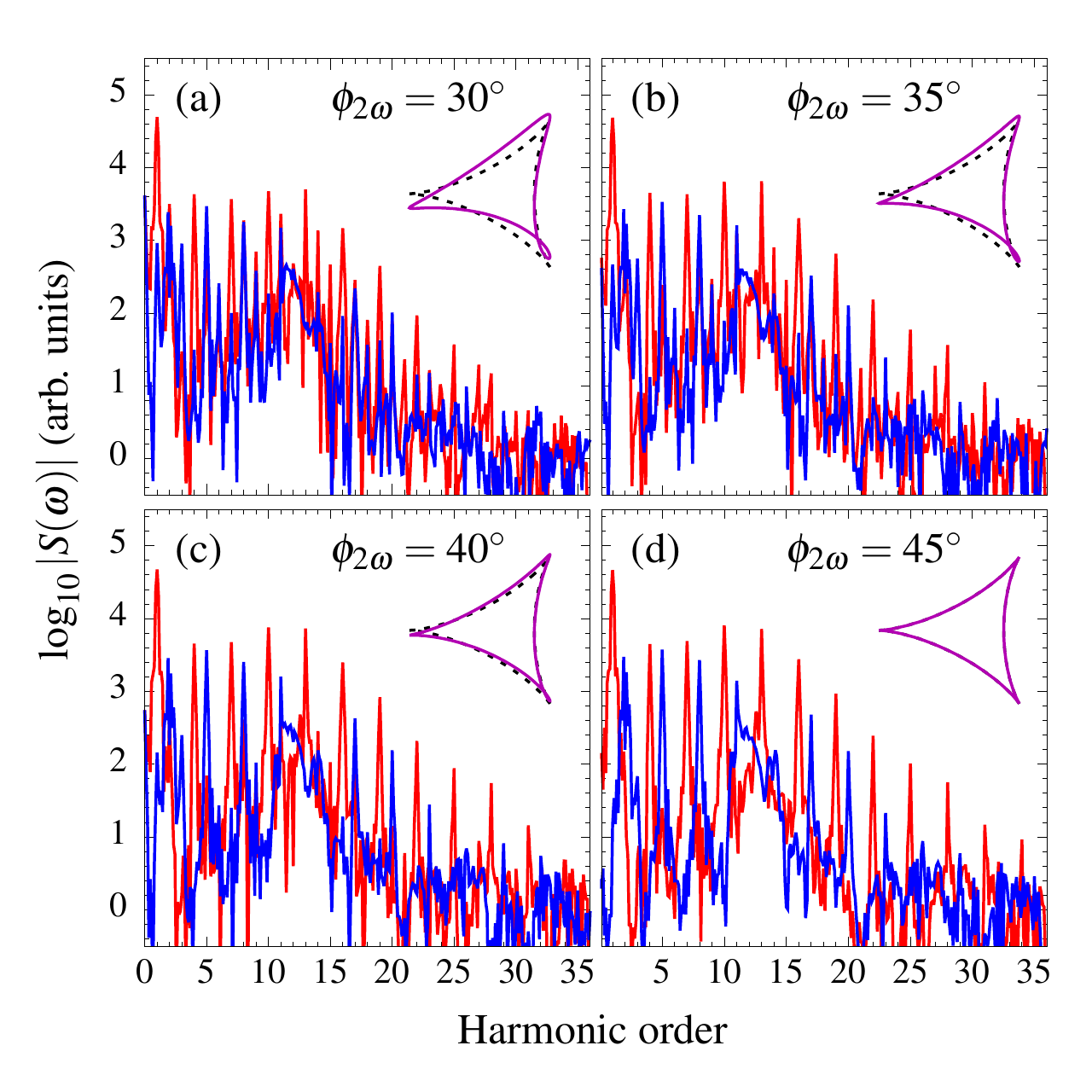}
	\caption{High-order harmonic spectrum of helium in $\omega - 2\omega$ driving fields of varying ellipticity. The $\omega$ field is kept perfect circular, i.e. $\phi_{\omega} = 45^{\circ}$, while the ellipticity of $2\omega$ driver is varied : (a) $\phi_{2\omega} = 30^{\circ}$, (b) $\phi_{2\omega} = 35^{\circ}$, (c) $\phi_{2\omega} = 40^{\circ}$, and (d) $\phi_{2\omega} = 45^{\circ}$. Rest of the pulse parameters are same as in Fig. \ref{hhg4545}(a). In the HHG spectrum, red lines indicate $D_+$ harmonic component, which co-rotates with $\omega$ field (counter-clockwise), while blue lines indicate $D_-$ harmonic components co-rotating with $2\omega$ field (clockwise). The insets show Lissajous curves of corresponding driving electric fields (purple solid line) for one optical cycle of the $\omega$ field. For comparison purpose, Lissajous curve of bicircular field (black dashed line) corresponding to $\phi_{\omega} = \phi_{2\omega} = 45^{\circ}$ is also shown.}
	\label{hhg_theta_vary}
\end{figure} 

\begin{figure*}[t]
	\includegraphics[totalheight=1.2\columnwidth]{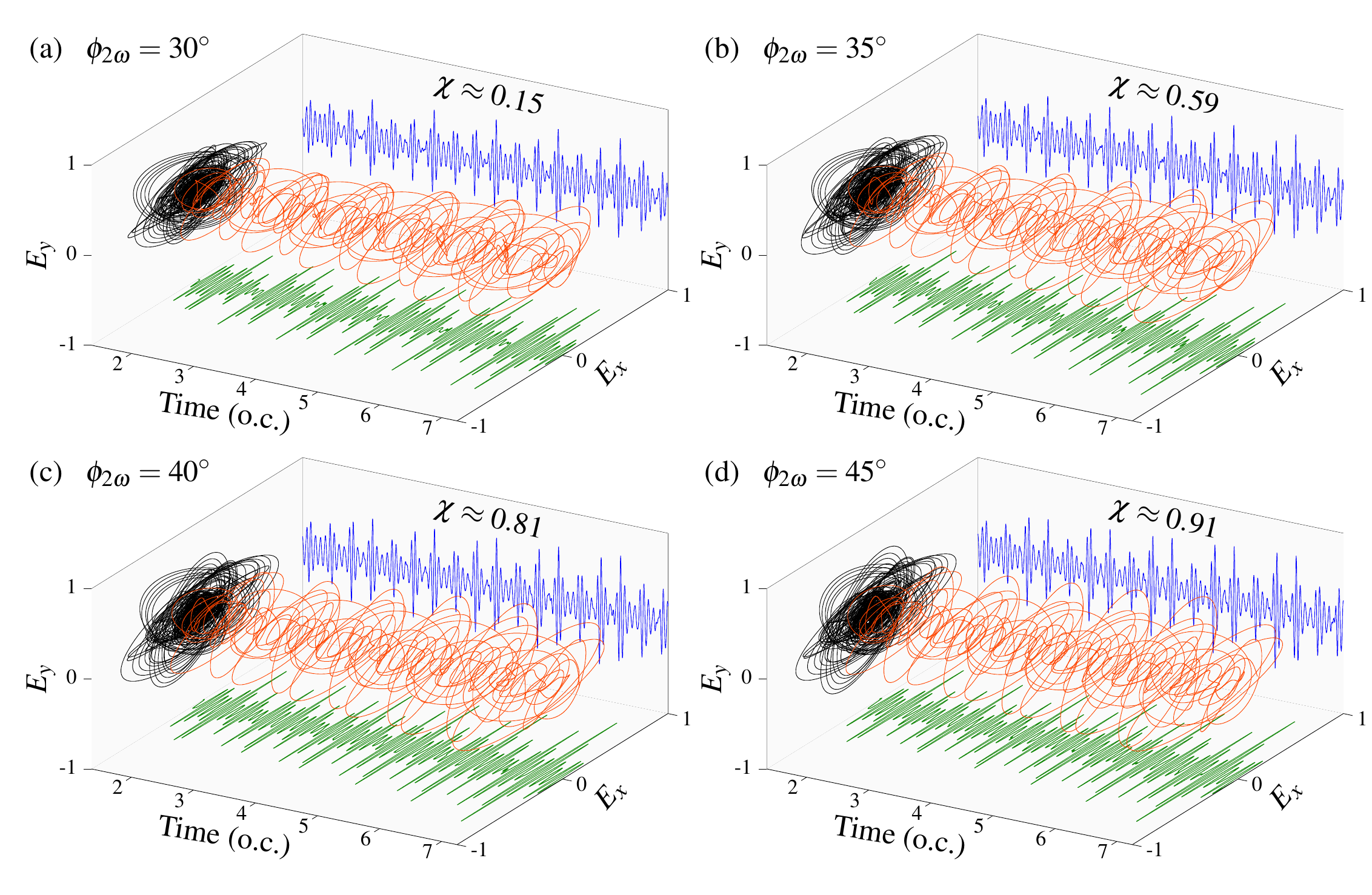}
	\caption{Attosecond pulse trains (APTs) generated by taking an energy window of 13th$-$29th harmonics for different rotation angle ($\phi_{2\omega}$) of 400 nm quarter-waveplate (a) $\phi_{2\omega} = 30^\circ$, (b) $\phi_{2\omega} = 35^\circ$, (c) $\phi_{2\omega} = 40^\circ$, and (d) $\phi_{2\omega} = 45^\circ$. The total time-dependent electric of APT is shown as orange curve. The $x$ component (green), $y$ component (blue), and Lissajous curve (black) are also presented. The ellipticity ($\chi$) of the generated APTs is given. The electric-field strengths $E_x$ and $E_y$ are given in arbitrary units.}
	\label{3dASP}
\end{figure*} 

We now discuss the effect of the ellipticity of $2\omega$ driver on the polarization properties of the generated harmonic spectrum. Fig. \ref{hhg_theta_vary} presents the HHG spectra of helium for different ellipticities of the second harmonic ($2\omega$) beam. The fundamental ($\omega$) beam is kept right (counter-clockwise) circularly polarized, that is, the rotation angle of $800$ nm quarter-waveplate is fixed at $\phi_{\omega} = 45^\circ$. The rotation angle of $400$ nm quarter-waveplate is scanned from $30^\circ$ to $45^\circ$ and the harmonic spectrum at $\phi_{2\omega} = 30^\circ, 35^\circ, 40^\circ,$ and $45^\circ$ are shown in Figs. \ref{hhg_theta_vary}(a)-(d), respectively. Besides the rotation angle $\phi_{2\omega}$, all the remaining pulse parameters are kept similar to the case shown in Fig. \ref{hhg4545}(a). In the HHG spectrum, red lines indicate the $D_+$ harmonic components, which co-rotate with the $\omega$ field (counter-clockwise), while the blue lines indicate $D_-$ harmonic component, which co-rotates with the $2\omega$ field (clockwise). From Fig. \ref{hhg_theta_vary}, we can see that the forbidden $3q\omega$ harmonics surfaced, and their intensity increases with the decrease in the ellipticity of the second-harmonic field. The appearance of $3q$ order harmonics is related to breaking the dynamical symmetry in the system. For $3q\omega$ harmonics, the blue lines dominate in the below-threshold energy harmonics (for He, threshold energy $I_p = 0.9$ a.u. $\approx 16\omega$), while the red lines dominate in the above-threshold energy harmonics  \cite{Galan2018_PRA, Heslar2017_PRA, Heslar2018_PRA}. It should also be noted in Fig. \ref{hhg_theta_vary} that the intensity of high harmonics generated for $\phi_{\omega} = 45^\circ$, $\phi_{2\omega} < 45^\circ$ cases, are comparable to the intensities of harmonics emitted for the case of bicircular field, i.e., $\phi_{\omega} = \phi_{2\omega} = 45^\circ$.
 
So far, we have discussed how the ellipticity of $2\omega$ driver affects the generated harmonic spectrum, specifically, the appearance of $3q$ order harmonics and a decrease in the circularity of individual harmonic peaks is observed. These effects translate into the ellipticity of generated attosecond pulses. In Figs. \ref{3dASP}(a)-(d), we show the total electric field (orange line) of the attosecond pulse train (APT) computed by taking the inverse Fourier transform of the corresponding harmonic spectra [refer Fig. \ref{hhg_theta_vary} and Eq. \ref{Ix}]. A band of harmonics from 13th to 29th order are filtered out for the construction of the pulse train. The $x$ component (green line), $y$ component (blue line), and the Lissajous curve (black line) are also shown. It can be seen that there are three XUV bursts per laser cycle ($T = 110.32$ a.u.) of the fundamental field. Due to the electric field amplitude ratio $E_1:E_2 = 2:1$ of the fundamental and the second harmonic beam, the bursts from the APT are highly elliptically polarized and co-rotate with the fundamental driver (counter-clockwise). The ellipticity of attosecond pulse train is calculated by integrating the two counter-rotating components $E_{\pm} = \mp (E_x \pm i E_y)/ \sqrt{2}$ of the total electric field over a time interval. The ellipticity is then defined as $\chi = (|E_+|^2 - |E_-|^2) / (|E_+|^2 + |E_-|^2)$ \cite{Galan2018_PRA}. The value of ellipticity $\chi$ varies from $-1$ to $+1$, corresponding to the two counter-rotating elliptically polarized ($|\chi| > 0$) electric fields. $\chi = 0$ and $|\chi| = 1$ correspond to the linear and perfect circular polarization states of the electric field, respectively. In Figs. \ref{3dASP}(a)-(d), the value of $\chi$ for the corresponding APT is also shown. The ellipticity is calculated for a temporal window of $150 - 800$ a.u. [same interval as shown in Fig. \ref{3dASP}]. It can be seen that the ellipticity of APT can be controlled by simply rotating one of the quarter-waveplates. The degree of polarization of generated APT increases with the increasing circularity of the $2\omega$ driver. According to the numerical calculation, for $\phi_{2\omega} = 30^\circ$ we obtain $\chi \approx 0.15$ (close to linear polarization). As $\phi_{2\omega}$ increase to $\phi_{2\omega} = 35^\circ, 40^\circ,$ and $45^\circ$, the ellipticity of the generated attosecond pulse train increase as $0.59, 0.81,$ and 0.91 (nearly circularly polarized), respectively.

\begin{figure}[t]
	\includegraphics[totalheight=0.65\columnwidth]{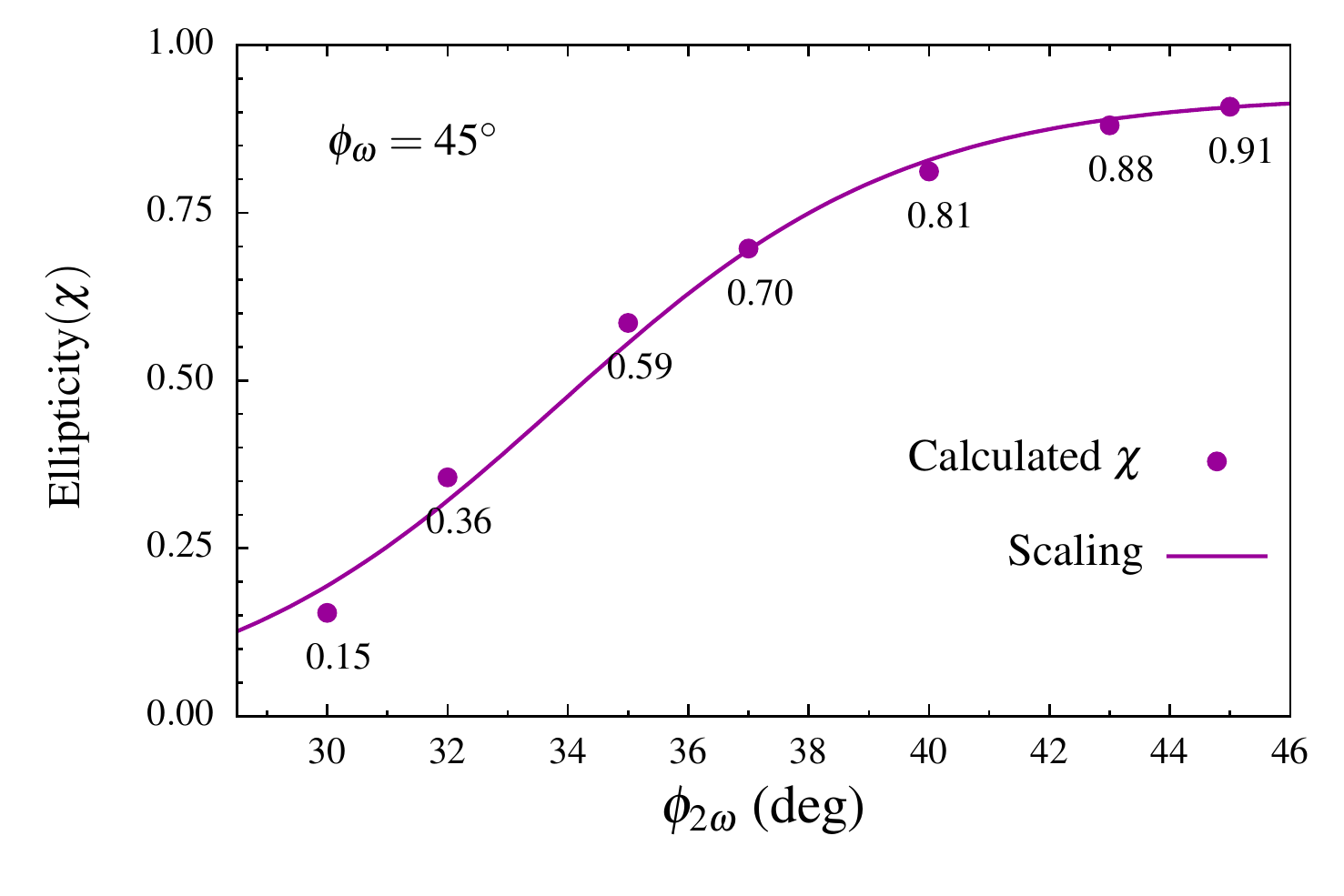}
	\caption{The calculated APT ellipticity $\chi$ (purple circle) for different rotation angle ($\phi_{2\omega}$) of the 400 nm quarter-waveplate. The calculated $\chi$ is labeled with the corresponding values. The fitting curve of the ellipticity scaling with $\phi_{2\omega}$ is depicted as purple solid line [refer to the text for more details]. }
	\label{ASPscaling}
\end{figure}  
 
In Fig. \ref{ASPscaling}, the calculated ellipticity ($\chi$) of attosecond pulse train for different rotation angles ($\phi_{2\omega}$) of 400 nm quarter-waveplate is presented. The ellipticity values correspond to the attosecond pulse trains shown in Fig. \ref{3dASP}. Besides those, ellipticity values at some intermediate angles $\phi_{2\omega} = 32^\circ, 37^\circ,$ and $43^\circ$ are also shown. The degree of polarization of generated APTs is critically sensitive to the ellipticity of the driving fields. The value of ellipticity increases smoothly along with the increasing circularity of the second-harmonic driving field. Based on these observations, we infer that the APT ellipticity in the current setup [refer Fig. \ref{fig_geo}] traces out the \emph{logistic curve}. The scaling of the ellipticity with angle $\phi_{2\omega}$ can be given as $\chi \propto 1/( 1 + \exp[-b(\phi_{2\omega} - m)] )$, where parameters $b$ and $m$ are fitting constants. This simple scaling provides the opportunity to fine-tune the polarization of generated ASPs and pulse train, all the way from linear through elliptical to circular, by simply changing the rotation angle of the quarter-waveplate.

It should be noted that the simple scaling presented here is for helium, i.e., the atomic system with spherically symmetric $s$ orbit valence electron. For the case of atoms with higher orbital angular momentum values (i.e., $l \geq 1$) or molecular systems with random orientations, the exact scaling is not obligatory. However, the harmonics emitted during the HHG process carry the signature of the path the electron progressed through during its excursion in the continuum. Thus, the ellipticity of driving fields greatly affects the polarization properties of the emitted harmonics irrespective of the generating medium.

\section{Concluding remarks}
\label{sec4}

In summary, we have theoretically investigated the HHG and the generation of attosecond pulses from helium using the bi-chromatic counter-rotating elliptically polarized driving fields. The dependence of harmonic polarization state on the laser pulse parameters offers an opportunity to shape the polarization properties of the emitted attosecond pulses. We have presented a simple scaling of APT ellipticity with the variation in the ellipticity of the driving electric field. This scaling provides the opportunity to fine-tune the ellipticity of generated attosecond pulses by simply changing the rotation angle of the quarter-waveplate [refer Fig. \ref{fig_geo}], which is quite feasible from an experimental point of view. This study helps us to generate attosecond pulses of varying degrees of polarization, all the way from linear to circular. These elliptically polarized APTs have a broad range of applications, such as ultrafast chiral recognition via photoelectron circular dichroism \cite{Bowering2001_PRL, Nahon2015_JElectronSpectrosc, Beaulieu2017_Science}, ultrafast XUV magnetization, and spin dynamics \cite{Turgut2013_PRL, boeglin2010_Nature, Kfir2017_SciAdv, Willems2015_PRB}. 
  
\section*{Acknowledgments} Authors would like to acknowledge the DST-SERB, Government of India, for funding the project CRG/2020/001020.

%\bibliographystyle{apsrev4-1}
%\bibliography{Bibliography}


\end{document}